\documentclass[a4paper]{article}
\usepackage{amsmath,amssymb,amsfonts,amsthm,amscd}
\usepackage[titletoc,title]{appendix}
\usepackage[bookmarks=true]{hyperref}
\usepackage{color}
\numberwithin{equation}{section}

\newcommand{\bea}{\begin{eqnarray}}
\newcommand{\eea}{\end{eqnarray}}
\newcommand{\be}{\begin{equation}}
\newcommand{\ee}{\end{equation}}
\newcommand{\nn}{\nonumber \\}

\begin{document}
\begin{titlepage}

\vfill \vfill \vfill
\begin{center}
{\bf\Large Non-linear ${\bf (3, 4, 1)}$ multiplet of ${\cal N}=4, d=1$ supersymmetry
\vspace{0.2cm}

as a semi-dynamical spin multiplet}
\end{center}
\vspace{1.5cm}

\begin{center}
{\large\bf Evgeny Ivanov${\,}^{a)\,b)}$, Stepan Sidorov${\,}^{a)}$}
\end{center}
\vspace{0.4cm}

\centerline{${\,}^{a)}$ \it Bogoliubov Laboratory of Theoretical Physics, JINR, 141980 Dubna, Moscow Region, Russia}
\vspace{0.2cm}
\centerline{${\,}^{b)}$ \it  Moscow Institute of Physics and Technology,
141700 Dolgoprudny, Moscow Region, Russia}
\vspace{0.3cm}

\centerline{\tt eivanov@theor.jinr.ru, sidorovstepan88@gmail.com}
\vspace{0.2cm}

\vspace{2cm}

\par

{\abstract
\noindent
We consider a new type of ${\cal N}=4$, $d=1$ semi-dynamical multiplet based on
the non-linear version of the mirror multiplet ${\bf (3, 4, 1)}$, with the triplet of bosonic physical fields parametrizing a three-dimensional sphere $S^3$ of the radius $R$.
The limit $R\to\infty$ amounts to the contraction $S^3\to\mathbb{R}^3$ and leads to the linear mirror multiplet ${\bf (3, 4, 1)}$.
Spin degrees of freedom described by a Wess-Zumino action specify a two-dimensional surface embedded in the sphere $S^3$.
A pair of the examples considered correspond to the round and squashed ``fuzzy'' 2-spheres.
We couple the squashed 2-sphere model to the dynamical mirror multiplet ${\bf (2, 4, 2)}$. A notable feature of this coupling is the dependence
of the squashing parameter on the bosonic fields $z, \bar z$ of the chiral multiplet.}
\vfill{}


\end{titlepage}
\section{Introduction}
Supersymmetric quantum mechanics (SQM) attracts permanent interest as a good training ground for studying various notorious features
of supersymmetric $d > 1$ field theory models \cite{Witt1,Witt2}.
An extended $d=1$ supersymmetry, including its superconformal enlargements (see, e.g., \cite{FILrev}), is of interest also on its own, as the relevant theories reveal
specific target geometries having no direct analogs in $d > 1$ systems (see, e.g., \cite{SmilgaBook} and refs therein).

The finite-component minimal off-shell multiplets used in constructing diverse models of ${\cal N}$-extended SQM are conveniently represented by triples ${\bf \left(k, \,\mathcal{N},\, \mathcal{N}-k\right)}$, where ${\bf k}$
stands for the number of physical bosonic fields, $\mathcal{N}$ for the number of fermionic fields and ${\bf \mathcal{N}-k}$ for the number of bosonic auxiliary fields \cite{PashTopp}. Apart from multiplets described by
$d=1$ superfields subjected to the constraints which are linear in the relevant covariant derivatives and so imply linear transformation rules for the component fields, there exist their non-linear cousins defined by the
proper non-linear superfield constraints (see, e.g., \cite{IKL2003} and \cite{DI2011}).  Generically, all these multiplets, both linear and non-linear, can further be divided into the two large classes depending on their
Lagrangian description: dynamical and semi-dynamical ones. At the classical $d=1$  Lagrangian level, the first type is characterized by the standard kinetic terms, of the second order in the time derivative for physical
bosonic fields and of the first order for fermionic fields. The second type implies $d=1$ Wess-Zumino (or Chern-Simons) term of the first order in the time derivative for bosonic field, while the fermionic components of the
``physical'' dimension do not appear with time derivatives at all and so must be treated as auxiliary fields (either as some Lagrange multipliers). The semi-dynamical multiplets are used to describe the spinning (or
``isospin'') degrees of freedom after quantization \cite{FIL1,FIL2,FIL3, IKS1,IKS2}\footnote{To avoid possible misunderstanding, we point out that the same supermultiplet can be dynamical or semi-dynamical for different
choices of the underlying off-shell action.}.

A specific feature of ${\cal N}=4$, $d=1$ supersymmetry is the existence of ``mirror'' (or ``twisted'') multiplets which differ from the standard ones by permuting
two automorphism ${\rm SU}(2)$ groups of ${\cal N}=4$, $d=1$ superalgebra. Both types of the multiplets have the same numbers of bosonic and fermionic components,
but their  transformation properties with respect to the automorphism groups turn out to be accordingly switched.  Recently \cite{IS2021}, we have considered couplings of the semi-dynamical
mirror linear multiplet ${\bf (3, 4, 1)}$ to the dynamical mirror linear multiplets. We reproduced the model constructed in \cite{FIL2012}, but in terms of
the harmonic mirror superfields. As a new result we presented the coupling of the semi-dynamical mirror multiplet ${\bf (3, 4, 1)}$ to the dynamical mirror multiplet ${\bf (2, 4, 2)}$
described by a chiral superfield, where the corresponding interaction term was constructed as a superpotential in chiral ${\cal N}=4, d=1$ superspace.
Here we generalize some of these results to the case of non-linear semi-dynamical mirror multiplet ${\bf (3, 4, 1)}$.

In Section 2 we describe the non-linear ${\bf (3, 4, 1)}$ multiplet in the standard ${\cal N}=4$, $d=1$ superspace and identify three mutually commuting ${\rm SU}(2)$ symmetry groups of its defining
superfield constraints. Two of them constitute ${\rm SO}(4)$ isometry of the $3$-sphere $S^3$ which is parametrized by the bosonic physical fields of the multiplet.
In Section 3 we construct the superfield off-shell Wess-Zumino (WZ) action for the non-linear multiplet, using the general action formula suggested earlier in \cite{IS2021}. As distinct from the linear ${\bf (3, 4, 1)}$
multiplet considered there, in the case under consideration the superfield Lagrangian density should satisfy the Laplace-Beltrami (LB) equation on the 3-sphere $S^3$. As instructive examples we  give the explicit form of the
solutions with maximal symmetries. These are superfield WZ actions giving rise, in the component approach, to the Fayet-Iliopoulos (FI) term (with the Lagrangian proportional to the bosonic auxiliary field and invariant
under all three ${\rm SU}(2)$ symmetries mentioned above) and to the Lorentz-forth-type coupling to the modified external Dirac monopole-type magnetic potential (with the product of ${\rm SU}(2)$ symmetries being broken to
some its subgroup). Integrating out the auxiliary bosonic field as the Lagrange multiplier we finally arrive at the second class Hamiltonian constraint which effectively eliminates one bosonic degree of freedom (resulting in
Dirac brackets for the rest of variables) and yields the two-sphere $S^2$ (or its some deformation) properly embedded in $S^3$ (in the case of linear ${\bf (3, 4, 1)}$ there appears an analogous two-sphere embedded in the flat $\mathbb{R}^3$ space
\cite{FIL2012,IS2021}). We also perform (at the component level) the ``oxidation'' procedure giving rise to the WZ action for a non-linear multiplet ${\bf (4,4,0)}$. In Section 3 we pass to the description of the non-linear
multiplet  ${\bf (3, 4, 1)}$ in terms of chiral ${\cal N}=4$, $d=1$ superfields, generalizing the similar formulation of the linear ${\bf (3, 4, 1)}$ multiplet in ref. \cite{IS2021}. Such a chiral setting enables a simple
construction of coupling of the non-linear multiplet to the dynamical chiral ${\bf (2, 4, 2)}$ multiplet. As a preliminary step, we  find the WZ Lagrangian which generalizes, to the non-linear case, the WZ Lagrangian
describing the coupling to the non-commutative 2-plane \cite{IS2021}.  Its non-linear ${\bf (3, 4, 1)}$ analog proves to be associated with a squashed two-sphere embedded in $S^3$, with the ``radius'' being related to the
strength of the FI term. After that we construct the WZ type interaction with the dynamical chiral multiplet ${\bf (2, 4, 2)}$ and find that the relevant Hamiltonian  second class constraint also gives rise to a sort of the
squashed $S^2$, but with a non-constant radius as a function of the physical bosonic fields of the chiral multiplet.

\section{Non-linear mirror multiplet ${\bf (3, 4, 1)}$}
Recall first the basic notions of ${\cal N}=4$, $d=1$ supersymmetric mechanics.
The standard ${\cal N}=4$, $d=1$ superalgebra reads
\bea
    \left\lbrace Q^{i}_{\alpha}, Q_{j}^{\beta}\right\rbrace = 2\delta^i_j\delta_\alpha^\beta H.
\eea
The ${\rm SU}(2)$ factors of the full automorphism group ${\rm SU}(2)_{\rm L} \times {\rm SU}(2)_{\rm R}$ acts on Latin ($i=1,2$) and Greek ($\alpha = 1,2$) indices, respectively.
The ${\cal N}=4$, $d=1$  superspace is parametrized by a time coordinate $t$ and four Grassmann coordinates $\theta^{i\alpha}$.
${\cal N}=4$ supersymmetry acts on these coordinates as
\bea
    \delta \theta^{i\alpha}= \epsilon^{i\alpha}, \qquad  \delta t = -\,i\,\epsilon^{i\alpha}\theta_{i\alpha}\,,
    \qquad \overline{\left(\theta^{i\alpha}\right)}=-\,\theta_{i\alpha}\,,\qquad \overline{\left(\epsilon^{i\alpha}\right)}
    =-\,\epsilon_{i\alpha}\,, \label{SStr}
\eea
where $\epsilon^{i\alpha}$ is a quartet of infinitesimal Grassmann parameters. The covariant derivatives $D^{i\alpha}$ are defined as
\bea
    &&D^{i\alpha}=\frac{\partial}{\partial \theta_{i\alpha}}+i\,\theta^{i\alpha}\partial_t\,,\qquad \overline{\left(D^{i\alpha}\right)} = D_{i\alpha}\,.
\eea

The non-linear mirror multiplet ${\bf (3, 4, 1)}$ is described by a quartet superfield $N^{\alpha A}$, where $A=1,2$ index corresponds
to some external group ${\rm SU}(2)_{\rm ext.}$ commuting with supersymmetry\footnote{We name this multiplet ``mirror'' since it is chosen to be a doublet of the automorphism
group ${\rm SU}(2)_{\rm R}$ which we associate with mirror ${\cal N}=4$  multiplets. We could equally choose it as a doublet of ${\rm SU}(2)_{\rm L}$ (with replacing the index $\alpha$ by $i$).
The convention we use here is more convenient for our purposes.}. This superfield satisfies the constraints \cite{IKL2003}\footnote{In the context of $4D$, ${\cal N}=2$
supersymmetry, this multiplet was considered for the first time in \cite{WHP}. Various aspects of the superfield description of its $d=1$ version were treated earlier in  refs. \cite{IKL2003}, \cite{BK2006}
and \cite{DeIv}.}:
\bea
    N^{\alpha A}N_{\alpha A} = R^2,\qquad N^{(\alpha}_{A}D^{\beta}_{i}N^{\gamma)A}=0, \qquad \overline{\left(N^{\alpha A}\right)}=N_{\alpha A}\,. \label{NL341quartet}
\eea
The first constraint specifies the 3-sphere $S^3$ in $\mathbb{R}^4$.
The constraints \eqref{NL341quartet} are covariant with respect to the total bosonic symmetry group ${\rm SU}(2)_{\rm L}\times{\rm SU}(2)_{\rm R}\times {\rm SU}(2)_{\rm ext.}$.
The subgroup ${\rm SU}(2)_{\rm R}\times {\rm SU}(2)_{\rm ext.}$ just forms ${\rm SO}(4)$ isometry group of $S^3$.

We introduce the stereographic coordinates $V^{\alpha\beta}$ on $S^3$ as
\bea
    V^{12}=\frac{R\left(N^{21}+N^{12}\right)}{\sqrt{2}\,R+N^{21}-N^{12}}\,,\qquad
    V^{11}=\frac{2RN^{11}}{\sqrt{2}\,R+N^{21}-N^{12}}\,,\qquad
    V^{22}=\frac{2RN^{22}}{\sqrt{2}\,R+N^{21}-N^{12}}\,,\label{FromVtoN}
\eea
where
\bea
    V^{\alpha\beta}=V^{\beta\alpha},\qquad \overline{\left(V^{\alpha\beta}\right)}=V_{\alpha\beta}\,.
\eea
The inverse mapping is
\bea
    &&N^{21}=\frac{-\,R\left(V^{2}-R^2-2R\,V^{12}\right)}{\sqrt{2}\left(R^2+V^{2}\right)}\,,\qquad
    N^{11}=\frac{\sqrt{2}\,R^2\,V^{11}}{R^2+V^{2}}\,,\nn
    &&N^{12}=\frac{R\left(V^{2}-R^2+2R\,V^{12}\right)}{\sqrt{2}\left(R^2+V^{2}\right)}\,,\qquad
    N^{22}=\frac{\sqrt{2}\,R^2\,V^{22}}{R^2+V^{2}}\,,\qquad V^{2}:=\frac{1}{2}\,V^{\alpha\beta}V_{\alpha\beta}\,.\label{inverse}
\eea
The conformally flat metric on the 3-sphere reads
\bea
    ds^2 = dv^{\alpha\beta}dv_{\alpha\beta}\left(1+\frac{v^2}{R^2}\right)^{-2},\qquad v^{\alpha\beta}:=V^{\alpha\beta}\left.\right|_{\theta=0}\,,\qquad
    v^{2}:=\frac{1}{2}\,v^{\alpha\beta}v_{\alpha\beta}\,. \label{metric}
\eea
In the limit $R\to\infty$ the flat $\mathbb{R}^3$ metric is recovered. More information on the geometry of the 3-sphere is adduced in Appendix \ref{AppA}.

The constraints \eqref{NL341quartet} rewritten in terms of the triplet $V^{\alpha\beta}$ take
the form
\bea
    D^{(\gamma}_{i}V^{\alpha\beta)}-\frac{1}{R}\,V^{\lambda(\gamma}D_{i\lambda}V^{\alpha\beta)}=0.\label{NL341}
\eea
One observes that in the limit $R\to\infty$  the linear constraints of the mirror multiplet ${\bf (3, 4, 1)}$ reappear \cite{IS2021}.
The non-linear constraints keep their form under the substitution $V^{\alpha\beta}\to -\,R^2\,V^{\alpha\beta}/V^2$ which relates
the $V^2<R^2$ and $V^2>R^2$ patches of the stereographic projection. For the quartet superfield \eqref{inverse}, it amounts just to the reflection $N^{\alpha A} \to -\,N^{\alpha A}$.

The non-linear constraints \eqref{NL341} are invariant under the following rotations and translations on the 3-sphere:
\bea
    &&\delta V^{\alpha\beta} = 2b^{(\alpha}_{\lambda} V^{\beta)\lambda},\quad\delta \theta^{i\alpha} = b^{\alpha}_{\lambda} \theta^{i\lambda},\quad \delta D^{\alpha}_{i} = b^{\alpha}_{\lambda} D^{\lambda}_{i},\qquad b_{\alpha\beta}=b_{\beta\alpha}\,,\quad\overline{\left(b_{\alpha\beta}\right)}=b^{\alpha\beta},\label{S3tr_diag}\\
    &&\delta V^{\alpha\beta} = a^{\alpha\beta} + \frac{2}{R}\,a^{(\alpha}_{\lambda}V^{\beta)\lambda} + \frac{1}{R^2}\,a^{\lambda\mu}V^{\alpha}_{\lambda}V^{\beta}_{\mu},\qquad a_{\alpha\beta}=a_{\beta\alpha}\,,\quad \overline{\left(a_{\alpha\beta}\right)}=a^{\alpha\beta}.\label{S3tr_ext}
\eea
The rotations with the parameters $b^{\alpha\beta}$ constitute the diagonal subgroup ${\rm SU}(2)_{\rm rot.}\subset {\rm SU}(2)_{\rm R}\times {\rm SU}(2)_{\rm ext.}$\,, while the non-linear
``translations'' with the parameters $a^{\alpha\beta}$ correspond to the external group ${\rm SU}(2)_{\rm ext.}$. The same transformations are linearly realized on $N^{\alpha A}$ as
\bea
    &&\delta N^{\alpha A} = b^{\alpha}_{\beta} N^{\beta A}+b^{\gamma}_{\lambda} \delta^A_\gamma \delta^\lambda_B N^{\alpha B},\qquad \delta D^{\alpha}_{i} = b^{\alpha}_{\beta} D^{\beta}_{i},\qquad
    \delta N^{\alpha A} = \frac{2}{R}\,a^{\gamma}_{\lambda} \delta^A_\gamma \delta^\lambda_B \,N^{\alpha B}.\label{quartet_tr}
\eea

In what follows we will be interested in construction of a superfield WZ action for the non-linear ${\bf (3, 4, 1)}$ multiplet. To this end,
we need to know only the first and second components in the $\theta$-expansion of  $V^{\alpha\beta}$.
The solution of \eqref{NL341} reads
\bea
    V^{\alpha\beta}&=&v^{\alpha\beta}+i\,\theta^{k(\alpha}\chi^{\beta)}_{k}-\frac{i}{R}\,\theta^{k}_{\lambda}v^{\lambda(\alpha}\chi^{\beta)}_{k}
    +i\left[\frac{1}{2}\,\theta^{k(\alpha}\theta^{\beta)}_{k}-\frac{1}{R}\,\theta^{k(\alpha}\theta_{k\gamma}v^{\beta)\gamma}
    +\frac{1}{2R^2}\,\theta^{k(\lambda}\theta_{k}^{\gamma)}v^{(\alpha}_{\lambda}v^{\beta)}_{\gamma}\right]C\nn
    &&+\,i\left[\theta^{k(\alpha}\theta_{k\gamma}\dot{v}^{\beta)\gamma}
    -\frac{1}{R}\left(\theta^{k}_{\gamma}\theta_{k\mu}v^{\gamma\mu}\dot{v}^{\alpha\beta}+2\,\theta^{k(\alpha}\theta_{k\gamma}\dot{v}^{\beta)\mu}v^{\gamma}_{\mu}\right)
    -\frac{1}{R^2}\,\theta^{k\gamma}\theta_{k\lambda}v^{\lambda}_{\mu}v^{(\alpha}_{\gamma}\dot{v}^{\beta)\mu}\right]\left(1 +\frac{v^2}{R^2}\right)^{-1}\nn
   &&-\,\frac{1}{2R}\,\theta^{k(\alpha}\theta_{k\gamma}\chi^{i\beta)}\chi_{i}^{\gamma}-\frac{1}{8R^2}\,\theta^{(i}_{\lambda}\theta^{k)\lambda}v^{\alpha\beta}\chi_{i\mu}\chi_{k}^{\mu}+\frac{1}{2R^2}\,\theta^{k(\lambda}\theta^{\mu)}_{k}\chi^{i}_{\lambda}\chi_{i}^{(\alpha}v^{\beta)}_{\mu}\nn
   &&+\,\frac{1}{8R^2}\,\theta^{k(\lambda}\theta^{\mu)}_{k}\,v^{\alpha\beta}\left(\chi^{i}_{\lambda}\chi_{i\mu}+\frac{2}{R}\,v^{\gamma}_{\lambda}\,\chi^{i}_{\gamma}\chi_{i\mu}+\frac{1}{R^2}\,v^{\gamma}_{\lambda}\,v^{\delta}_{\mu}\,\chi^{i}_{\gamma}\chi_{i\delta}\right)\left(1 +\frac{v^2}{R^2}\right)^{-1}\nn
   &&+\left(\theta^3\;{\rm and}\;\theta^4\;{\rm terms}\right),
    \label{V}
\eea
where
\bea
    \overline{\left(v^{\alpha\beta}\right)}=v_{\alpha\beta}\,,\qquad v^{2}=\frac{1}{2}\,v^{\alpha\beta}v_{\alpha\beta}\,,\qquad \overline{\left(\chi^{k\alpha}\right)}
    =-\,\chi_{k\alpha}\,,\qquad \overline{\left(C\right)}=C.
\eea
The components transform as
\bea
    \delta_\epsilon v^{\alpha\beta} &=& -\,i\left(\epsilon^{k(\alpha}\chi^{\beta)}_{k}-\frac{1}{R}\,\epsilon^{k}_{\lambda}v^{\lambda(\alpha}\chi^{\beta)}_{k}\right),\nn
    \delta_\epsilon \chi^{\alpha}_{i} &=& -\,2\,\epsilon_{i\beta}\left(\dot{v}^{\alpha\beta}-\frac{1}{R}\,v^{\beta}_{\gamma}\dot{v}^{\alpha\gamma}\right)\left(1 +\frac{v^2}{R^2}\right)^{-1}
    -\left(\epsilon^{\alpha}_{i}-\frac{1}{R}\,\epsilon_{i\beta}v^{\alpha\beta}\right)C+\frac{i}{2R}\,\epsilon^{k\beta}\chi_{i\beta}\chi^{\alpha}_{k}\nn
    &&-\,\frac{i}{2R}\left[\epsilon_{i\beta}\left(\chi^{k\beta}-\frac{1}{R}\,v^{\beta}_{\gamma}\chi^{k\gamma}\right)\chi^{\alpha}_{k}-\frac{v_{\lambda\mu}}{2R}\,\left(\epsilon^{\alpha}_{i}-\frac{1}{R}\,\epsilon_{i\beta}v^{\alpha\beta}\right)\chi^{i\lambda}\chi_{i}^{\mu}\right]\left(1 +\frac{v^2}{R^2}\right)^{-1},\nn
    \delta_\epsilon C &=& -\,i\,\epsilon_{i\alpha}\,\partial_t\left[\left(\chi^{i\alpha}
    -\frac{1}{R}\,\chi^{i\beta}v^{\alpha}_{\beta}\right)\left(1 +\frac{v^2}{R^2}\right)^{-1}\right].\label{trcomp}
\eea

\subsection{Wess-Zumino Lagrangians}
The sigma-model type and WZ Lagrangians for the non-linear multiplet ${\bf (3, 4, 1)}$ were constructed in \cite{BK2006} using some ``joint-mixed'' formalism.
Only the component form of WZ Lagrangians (that were dubbed ``generalized Fayet-Iliopoulos terms''), was given there. Here, we consider WZ Lagrangian on its own and present
it in a manifestly ${\cal N}=4$ supersymmetric superfield form.
Such a description is most appropriate for our ultimate purpose to make use of the non-linear multiplet to represent semi-dynamical spin variables.

Based upon the paper \cite{IS2021}, we suggest the following  ansatz for the sought WZ action
\bea
    S_{\rm WZ} = \int dt\,{\cal L}_{\rm WZ} =
    \int dt\,d^4\theta\left[i\,\theta^{k(\alpha}\theta^{\beta)}_{k}\,L_{\alpha\beta}\left(V\right)\right],\qquad \overline{\left(L_{\alpha\beta}\right)}=L^{\alpha\beta},\quad L^{\alpha\beta}=L^{\beta\alpha},\label{WZaction}
\eea
where the triplet superfield function $L^{\alpha\beta}$ satisfies a quadratic condition introduced in \cite{IS2021}:
\bea
    D^{(i}_{\gamma}D^{j)\gamma}L^{\alpha\beta}\left(V\right)=0.\label{quadra}
\eea
Rewriting the $\theta$-integration measure as
\bea
    d^4\theta = \frac{1}{12}\,D_{i\lambda}D^{\lambda}_{j}D^{(i}_{\mu}D^{j)\mu},\label{D4}
\eea
we prove that under the supersymmetry transformations \eqref{SStr} the variation of \eqref{WZaction} is reduced to the expression which vanishes as a consequence of \eqref{quadra}:
\bea
    \delta_{\epsilon} S_{\rm WZ} = \frac{i}{6}\int dt\,D_{i\lambda}D^{\lambda}_{j}D^{(i}_{\mu}D^{j)\mu}
    \left[\epsilon^{k(\alpha}\theta^{\beta)}_{k}\,L_{\alpha\beta}\left(V\right)\right]=\frac{2i}{3}\int dt\,\epsilon_{j\alpha}\,D_{i\beta}D^{(i}_{\mu}D^{j)\mu}L^{\alpha\beta}\left(V\right)=0.\label{WZcond}
\eea
Exploiting the non-linear constraints \eqref{NL341}, we rewrite the quadratic condition \eqref{quadra} as
\bea
    \frac{1}{3}\,\left(D^{(i}_{\gamma}V_{\delta\rho}\right)\left(D^{j)\gamma}V^{\delta\rho}\right)
    \left[\frac{\partial}{\partial V_{\lambda\mu}}- \frac{V^{\lambda\mu}}{R^2}\left(1 +\frac{V^2}{R^2}\right)^{-1}\right]\frac{\partial}{\partial V^{\lambda\mu}}\,L^{\alpha\beta}\left(V\right)=0,\label{LBSFeq}
\eea
which suggests that the triplet function $L^{\alpha\beta}\left(v\right)$ satisfies the Laplace-Beltrami equation on $S^3$:
\bea
     {\bf \Delta}_{S^3}\,L^{\alpha\beta}\left(v\right)=0,\quad
     {\bf \Delta}_{S^3}= \left(1+\frac{v^2}{R^2}\right)^2\left[\partial^{\lambda\mu}- \frac{v^{\lambda\mu}}{R^2}\left(1 +\frac{v^2}{R^2}\right)^{-1}\right]\partial_{\lambda\mu}\,,\quad
     \partial^{\lambda\mu}v_{\alpha\beta}=\delta_{(\alpha}^{\lambda}\delta_{\beta)}^{\mu}.\label{LBeq}
\eea

In components, we arrive at the WZ Lagrangian in the form
\bea
    {\cal L}_{\rm WZ}=C\,{\cal U}\left(v\right)-\dot{v}^{\alpha\beta}{\cal A}_{\alpha\beta}\left(v\right)-\frac{i}{2}\,
    \chi^{i\alpha}\chi_{i}^{\beta}\,{\cal R}_{\alpha\beta}\left(v\right),\label{WZL}
\eea
where
\bea
    {\cal U}\left(v\right) &=& \partial_{\alpha\beta}L^{\alpha\beta}+\frac{2}{R}\,v^{\gamma}_{\alpha}\,\partial_{\beta\gamma}L^{\alpha\beta}
    +\frac{1}{R^2}\,v^{\gamma}_{\alpha}\,v^{\lambda}_{\beta}\,\partial_{\gamma\lambda}L^{\alpha\beta},\nn
    {\cal A}_{\alpha\beta}\left(v\right) &=& \left(-\,2\,\partial_{\gamma(\alpha}L^{\gamma}_{\beta)}-
    \frac{4}{R}\,v_{\gamma(\alpha}\,\partial_{\beta)\lambda}L^{\gamma\lambda}+ \frac{2}{R}\,v_{\gamma\lambda}\,\partial_{\alpha\beta}L^{\gamma\lambda}-\frac{2}{R^2}v_{\lambda}^{\rho}\,v_{\gamma(\alpha}\,\partial_{\beta)\rho}L^{\gamma\lambda}\right)\left(1 +\frac{v^2}{R^2}\right)^{-1},\nn
    {\cal R}_{\alpha\beta}\left(v\right) &=& \partial_{\mu(\alpha}\,\partial_{\beta)\lambda}L^{\lambda\mu}-\frac{2}{R}\left(\partial_{\gamma(\alpha}L^{\gamma}_{\beta)}-v^{\gamma}_{\lambda}\,\partial_{\mu(\alpha}\,\partial_{\beta)\gamma}L^{\lambda\mu}\right)\nn
    &&-\,\frac{1}{R^2}\left(2\,v^{\lambda\mu}\,\partial_{\lambda(\alpha}L_{\beta)\mu}+v^{\gamma}_{\lambda}v^{\rho}_{\mu}\,\partial_{\rho(\alpha}\,\partial_{\beta)\gamma}L^{\lambda\mu}\right)\nn
    &&+\,\frac{1}{2R^2}\left(v^{\lambda\mu}\,\partial_{\lambda\mu}L_{\alpha\beta}-\frac{2}{R}\,v^{\gamma}_{\alpha}\,v^{\lambda\mu}\,\partial_{\lambda\mu}L_{\gamma\beta}
    +\frac{1}{R^2}\,v^{\gamma}_{\alpha}\,v^{\delta}_{\beta}\,v^{\lambda\mu}\,\partial_{\lambda\mu}L_{\gamma\delta}\right)\left(1 +\frac{v^2}{R^2}\right)^{-1}.\label{UAR}
\eea
From these definitions and eqs. \eqref{LBeq} we derive that
\bea
    \partial^{\gamma}_{(\alpha}\,{\cal A}_{\beta)\gamma}=\partial_{\alpha\beta}\,{\cal U}\left(1 +\frac{v^2}{R^2}\right)^{-1},\qquad {\cal R}_{\alpha\beta}=\partial_{\alpha\beta}\,{\cal U}.
    \label{UAReqs}
\eea
These equations guarantee the invariance of ${\cal L}_{\rm WZ}$ under the component transformations \eqref{trcomp}.
Both equations can be rewritten as\footnote{The explicit expressions for the divergence, rotor, gradient and Laplace operators in terms of covariant derivatives are presented in Appendix \ref{AppA}.}
\bea
    {\rm\bf rot}_{S^3}\,{\cal A} = {\rm\bf grad}_{S^3}\,{\cal U},\qquad {\cal R}_{\alpha\beta}= \nabla_{\alpha\beta}\,{\cal U}. \label{rotA}
\eea
One can easily prove that the scalar ${\cal U}$ satisfies the Laplace-Beltrami equation \eqref{LBeq}:
\bea
    {\bf \Delta}_{S^3}\,{\cal U}=0.\label{LBU}
\eea
Generally, the vector potential ${\cal A}_{\alpha\beta}$ is defined up to the gauge freedom
${\cal A}_{\alpha\beta} \rightarrow {\cal A}_{\alpha\beta} + \partial_{\alpha\beta}\,f$.
However, the definition of the potential as in \eqref{UAR} implies the gauge-fixing condition
\bea
\left(1 +\frac{v^2}{R^2}\right)^2\left[\partial^{\alpha\beta}-\frac{v^{\alpha\beta}}{R^2}\left(1 +\frac{v^2}{R^2}\right)^{-1}\right]{\cal A}_{\alpha\beta} = -\,\frac{4\,{\cal U}}{R}\quad\Rightarrow\quad
{\rm\bf div}_{S^3}\,{\cal A} =  -\,\frac{4\,{\cal U}}{R}\,.\label{divA}
\eea

The WZ Lagrangian \eqref{WZL} is manifestly invariant under the ${\rm SU}(2)_{\rm L}$ group symmetry acting on the Latin index
of the fermionic field $\chi^{i\alpha}$.
The ${\rm SO}(4)\sim {\rm SU}(2)_{\rm R}\times {\rm SU}(2)_{\rm ext.}$ isometry of $S^3$ is broken, but a fraction of it can still be preserved.
According to the interpretation suggested in \cite{FIL2012}, the auxiliary field $C$ in \eqref{WZL} plays the role of a Lagrange multiplier.
Its equation of motion enforces the constraint ${\cal U}\left(v\right)\approx 0$ that kills one degree of freedom in the triplet $v^{\alpha\beta}$.
As a result, the original group ${\rm SO}(4)$ reduces to the isometry group of a 2-dimensional surface embedded in $S^3$.

The Fayet-Iliopoulos (FI) Lagrangian corresponds to a trivial constant solution of \eqref{LBU}. It is written, up to a total time derivative and an arbitrary renormalization constant, as
\bea
    {\cal L}_{\rm FI} = C \, \gamma,   \qquad \gamma ={\rm const.}\label{FIL}
\eea
The relevant off-shell superfield FI action in the form \eqref{WZaction} can be written as
\bea
    S_{\rm FI} = \int dt\,{\cal L}_{\rm FI} =
    \gamma\int dt\,d^4\theta\left[i\,\theta^{k\alpha}\theta^{\beta}_{k}\,L^{\rm FI}_{\alpha\beta}\left(V\right)\right], \label{LFI}
\eea
where the triplet $L^{\rm FI}_{\alpha\beta}$ satisfying \eqref{quadra} is given by
\bea
    &&L^{\rm FI}_{11}=\frac{V_{11}}{6}\left(1 + \frac{2V_{12}}{R} - \frac{V^2}{R^2}\right)^{-1},\qquad L^{\rm FI}_{22}=\frac{V_{22}}{6}\left(1 -  \frac{2V_{12}}{R} - \frac{V^2}{R^2} \right)^{-1},\nn
    &&L^{\rm FI}_{12}=\frac{R}{24}\left[\log{\left(1 + \frac{2V_{12}}{R} - \frac{V^2}{R^2}\right)}-\log{\left(1 - \frac{2V_{12}}{R} - \frac{V^2}{R^2}\right)}\right],\label{FIaction}
\eea
with $V^2 = \frac{1}{2}\, V^{\alpha\beta}V_{\alpha\beta}$\,. It can be checked to be invariant under the full isometry group ${\rm SO}(4)$ given by the transformations \eqref{S3tr_diag}-\eqref{S3tr_ext}\footnote{This invariance
becomes manifest \cite{SidIvN} in the $d=1$ harmonic superspace formulation \cite{HSSd1}.}.

Obviously, it makes no sense to consider the FI term alone, it is meaningful only in combination with other invariant Lagrangians, e.g. with the Lagrangian
\eqref{WZL} involving non-trivial non-zero potentials. In the presence of FI term the constraint ${\cal U}\left(v\right)\approx 0$ should
be replaced by ${\cal U}\left(v\right)+\gamma\approx 0$.

\subsection{Dirac brackets}\label{DB}
First we eliminate the fermionic fields in \eqref{WZL} by their equations of motion.
Then we pass to the Hamiltonian system with $\lambda^{\alpha\beta}$ and $C$ treated as Lagrange multipliers:
\bea
    H=\lambda^{\alpha\beta}\pi_{\alpha\beta}-C\,{\cal U}. \label{Ham}
\eea
The second class Hamiltonian constraints of the system (at $c=0$ for simplicity) are then given by
\bea
    \pi_{\alpha\beta}=p_{\alpha\beta}+{\cal A}_{\alpha\beta}\approx 0,\qquad  {\cal U}\approx 0. \label{pi}
\eea
The $4\times 4$ matrix formed by Poisson brackets of the constraints \eqref{pi} can be divided into the blocks ($3\times 3$, $3\times 1$, $1\times 3$, $1\times 1$) and  written as
\bea
M=\left(1+ \frac{v^2}{R^2}\right)^{-1}\begin{pmatrix}
-\,2\delta^{(\alpha}_{(\lambda}\partial^{\beta)}_{\mu)}\,{\cal U}  & -\left(1+\displaystyle\frac{ v^2}{R^2}\right)\partial^{\alpha\beta}\,{\cal U}  \\
\left(1+\displaystyle\frac{v^2}{R^2}\right)\partial_{\lambda\mu}\,{\cal U} &  0\\
\end{pmatrix},
\label{matrix}
\eea
where we made use of the relations \eqref{UAReqs}.
Its determinant can be calculated to be
\bea
\det{M}=\left(\partial_{\alpha\beta}\,{\cal U}\,\partial^{\alpha\beta}\,{\cal U}\right)^2\left(1+ \frac{v^2}{R^2}\right)^{-2}.\label{det}
\eea
Consequently, the non-degenerate matrix $M$ satisfies the condition
\bea
    \partial_{\alpha\beta}\,{\cal U}\,\partial^{\alpha\beta}\,{\cal U} \neq 0.
\eea
The inverse matrix is
\bea
M^{-1}=\frac{1}{\partial_{\gamma\delta}\,{\cal U}\,\partial^{\gamma\delta}{\cal U}}
\begin{pmatrix}
\left(1+\displaystyle\frac{v^2}{R^2}\right)\delta^{(\alpha}_{(\lambda}\partial^{\beta)}_{\mu)}\,{\cal U} &  \partial^{\alpha\beta}\,{\cal U}  \\
-\,\partial_{\lambda\mu}\,{\cal U}& 0 \\
\end{pmatrix}
\eea
The Dirac brackets are then found to be
\bea
    \left\lbrace v^{\alpha\beta}, v_{\lambda\mu}\right\rbrace_{\rm DB}
    = \frac{\delta^{(\alpha}_{(\lambda}\partial^{\beta)}_{\mu)}\,{\cal U}}{\partial_{\gamma\delta}\,{\cal U}\,\partial^{\gamma\delta}\,{\cal U}}\left(1+\frac{v^2}{R^2}\right).
\eea

\subsection{Sphere $S^2$}
The monopole solution (or a ''fuzzy sphere $S^2$``) for the linear multiplet ${\bf (3, 4, 1)}$ was constructed in the analytic harmonic superspace \cite{HSSd1,FIL2012}.
Counterpart of this solution for the non-linear mirror multiplet ${\bf (3, 4, 1)}$ that satisfies \eqref{quadra}
is given by
\bea
    L^{{\rm SU}(2)}_{\alpha\beta}=\frac{2k_{\alpha\beta}V^{2}-k^{\lambda\mu}V_{\lambda\mu}V_{\alpha\beta}}{2|V|\left(2|V|+k^{\lambda\mu}V_{\lambda\mu}\right)}\,,
    \qquad k^2=\frac{1}{2}\,k^{\alpha\beta}k_{\alpha\beta}\equiv 1.\label{SU2}
\eea
The corresponding action \eqref{WZaction} is invariant under the rotations \eqref{S3tr_diag}.
The superfield function $L^{{\rm SU}(2)}_{\alpha\beta}$ bears no the explicit dependence on the parameter $R$, which is however hidden in the component expansion \eqref{V} of $V^{\alpha\beta}$.

The scalar and vector potentials are calculated according to the generic prescriptions \eqref{UAR}
\bea
    {\cal U}\left(v\right)=-\,\frac{1}{2|v|}\left(1-\frac{v^2}{R^2}\right),\qquad
    {\cal A}_{\alpha\beta} = \frac{k^{\gamma}_{(\alpha}v_{\beta)\gamma}}{|v|\left(2|v|+k^{\lambda\mu}v_{\lambda\mu}\right)}+\frac{v_{\alpha\beta}}{R\,|v|}\left(1+\frac{v^2}{R^2}\right)^{-1}.\label{UA}
\eea
Then we impose the constraint ${\cal U}\approx 0$ and derive the equation of the sphere $S^2$:
\bea
    {\cal U}\approx 0 \quad \Rightarrow \quad v^2\approx R^2. \label{Constr1}
\eea
For clarification, it is convenient to pass to coordinates on $\mathbb{R}^4$ related to the original quartet of fields $n^{\alpha A}:=N^{\alpha A}\left.\right|_{\theta=0}$ :
\bea
    &&n^{11}= \frac{1}{\sqrt{2}}\left(x_{1}+ix_{2}\right),\qquad
    n^{22}= \frac{1}{\sqrt{2}}\left(x_{1}-ix_{2}\right),\nn
    &&n^{12}= \frac{1}{\sqrt{2}}\left(x_{3}+ix_{4}\right),\qquad
    n^{21}= -\,\frac{1}{\sqrt{2}}\left(x_{3}-ix_{4}\right).\label{x4}
\eea
They satisfy the 3-sphere equation
\bea
    n^{\alpha A}n_{\alpha A}=\left(x_{1}\right)^2+\left(x_{2}\right)^2+\left(x_{3}\right)^2+\left(x_{4}\right)^2=R^2\label{3sphere}
\eea
(recall the first equation in \eqref{NL341quartet}). Exploiting the relation between the fields $v^{\alpha\beta}$ and $n^{\alpha A}$ implied by eqs. \eqref{FromVtoN} and \eqref{inverse} we can bring
the  constraint \eqref{Constr1} to the form
\bea
    {\cal U}\left(x_3\right)=\frac{x_3}{R\left[R^2-\left(x_{3}\right)^2\right]^{\frac{1}{2}}}\approx 0.\label{Ux3}
\eea
Thus, the sphere $S^2$ appears as an intersection of $S^3$ and the plane $x_3=0$.

One can modify the constraint \eqref{Ux3} by adding the FI-term \eqref{LFI} as
\bea
    {\cal U}\left(x_3\right)+\gamma=\frac{x_3}{R\left[R^2-\left(x_{3}\right)^2\right]^{\frac{1}{2}}}+\gamma\approx 0.
\eea
It preserves the rotation symmetry \eqref{S3tr_diag} and leads to the solution
\bea
    \left(x_3\right)^2\approx \gamma^2 R^4\left(1+\gamma^2 R^2\right)^{-1} \quad \Rightarrow \quad v^2\approx R^2\left( \sqrt{1+\gamma^2 R^2}-\gamma\,R\right)^2 := r^2.
\eea
We once again obtain the 2-sphere $S^2$, but with the radius $r$ specified by the FI constant $\gamma$:
\bea
    \gamma = \frac{1}{2r}\left(1-\frac{r^2}{R^2}\right),\qquad v^2\approx r^2.
\eea
The relevant embedding constraint is expressed as
\bea
    {\cal U}\left(v\right)+\gamma=\frac{1}{2r}\left(1-\frac{r^2}{R^2}\right)-\frac{1}{2|v|}\left(1-\frac{v^2}{R^2}\right)\approx 0 \label{S2inS3}
\eea
and it defines the sphere $S^2$ embedded in $S^3$. In the limit $R\to\infty$, we recover the sphere $S^2$ embedded in $\mathbb{R}^3$:
\bea
(2.47) \quad \Rightarrow \quad  {\cal U}\left(v\right)+\gamma=\frac{1}{2r}-\frac{1}{2|v|}\approx 0.\label{S2inR3}
\eea
The triplet $v^{\alpha\beta}$ satisfies the same Dirac brackets for both linear and non-linear cases:
\bea
    \left\lbrace v^{\alpha\beta}, v_{\lambda\mu}\right\rbrace_{\rm DB} = 2\,\delta^{(\alpha}_{(\lambda}\,v^{\beta)}_{\mu)}\,|v|.
\eea
One can check that they form the $su(2)$ algebra for which the square $v^2=r^2$ is Casimir operator.
So both cases, linear and non-linear, yield fuzzy 2-spheres, though embedded in $S^3$ and $\mathbb{R}^3$.
This latter difference implies that the relevant semi-dynamical  multiplets can have different interactions with dynamical multiplets.

\subsection{Oxidation to non-linear multiplet ${\bf (4, 4, 0)}$ and Nahm-type equations}
One observes that the auxiliary $C$ transforms as a full time derivative \eqref{trcomp}. This means that we can perform the ``oxidation'' procedure
as an inversion of the ``reduction'' procedure \cite{DI2006}. We substitute the auxiliary field as $C=\dot{x}$ and obtain the non-linear mirror multiplet ${\bf (4,4,0)}$ \cite{DI2011},
the bosonic part of which parametrizes the manifold $S^{3}\times \mathbb{R}^1$.

The WZ Langrangian \eqref{WZL} becomes
\bea
    {\cal L}_{\rm WZ}=\dot{x}\,{\cal U}\left(v\right)-\dot{v}^{\alpha\beta}{\cal A}_{\alpha\beta}\left(v\right)-\frac{i}{2}\,\chi^{i\alpha}\chi_{i}^{\beta}\,{\cal R}_{\alpha\beta}\left(v\right).
\eea
The Hamiltonian \eqref{Ham} is modified as
\bea
    H=\lambda^{\alpha\beta}\pi_{\alpha\beta}+\lambda\,\pi_x\,,\label{H440}
\eea
where the second class constraints are
\bea
    \pi_{\alpha\beta}=p_{\alpha\beta}+{\cal A}_{\alpha\beta}\approx 0,\qquad  -\,\pi_x = {\cal U}-p_x\approx 0.
\eea
They form the same matrix \eqref{matrix}, since  the dependence on $x$ is nowhere present.
Hence Dirac brackets are
\bea
    \left\lbrace v^{\alpha\beta}, v_{\lambda\mu}\right\rbrace_{\rm DB} = \frac{\delta^{(\alpha}_{(\lambda}\partial^{\beta)}_{\mu)}{\cal U}}{\partial_{\gamma\delta}\,{\cal U}\,\partial^{\gamma\delta}\,{\cal U}}\left(1+\frac{v^2}{R^2}\right),\qquad \left\lbrace v^{\alpha\beta}, x\right\rbrace_{\rm DB} =  -\,\frac{\partial^{\alpha\beta}{\cal U}}{\partial_{\gamma\delta}\,{\cal U}\,\partial^{\gamma\delta}\,{\cal U}}\,.\label{DB440}
\eea
The consideration can be simplified by the following trick. We rename the field $x$ as $x=:-\,p_{\tau}$ ($p_{x}=:\tau$) and obtain
 \bea
    &&\left\lbrace v^{\alpha\beta}, v_{\lambda\mu}\right\rbrace_{\rm DB} = \frac{\delta^{(\alpha}_{(\lambda}\partial^{\beta)}_{\mu)}{\cal U}}{\partial_{\gamma\delta}\,{\cal U}\,\partial^{\gamma\delta}\,{\cal U}}\left(1+\frac{v^2}{R^2}\right),\qquad \left\lbrace v^{\alpha\beta}, p_{\tau}\right\rbrace_{\rm DB} =  \frac{\partial^{\alpha\beta}{\cal U}}{\partial_{\gamma\delta}\,{\cal U}\,\partial^{\gamma\delta}\,{\cal U}}\,,\nn
    &&\left\lbrace \tau, p_{\tau}\right\rbrace_{\rm DB}=1,\qquad \left\lbrace \tau, v^{\alpha\beta}\right\rbrace_{\rm DB}=0.
\eea
We still deal with four independent variables: two represented by $\tau$ and $p_{\tau}$ and two hidden in the triplet $v^{\alpha\beta}$ due to the constraint $\tau\approx {\cal U}\left(v\right)$.
In the limit $R\to \infty$, the above brackets coincide with those found for the coupling of the dynamical multiplet ${\bf (1, 4, 3)}$ to the spin linear multiplet ${\bf (3, 4, 1)}$
(eqs. (3.18)-(3.20) in ref. \cite{FIL2012}). It is worth to note that the coupled system of \cite{FIL2012} has four bosonic and four fermionic degrees of freedom.
In the case under consideration, the fermionic fields $\chi^{i\alpha}$ are eliminated by their equations of motion.

Finally, rewriting the triplet as $v^{\alpha\beta}\rightarrow v_{a}$ ($a=1,2,3$), we derive the modified analog of the Nahm-type equations on $S^{3}\times \mathbb{R}^1$ :
\bea
    \partial_{\tau}v_{a}=\frac{1}{2}\,\varepsilon_{abc}\left\lbrace v_{b}, v_{c}\right\rbrace_{\rm DB}\left(1+\frac{v^2}{R^2}\right)^{-1}.
\eea
The $R\to \infty$ version of these equations was deduced in \cite{FIL2012} for a linear semi-dynamical multiplet $({\bf 3, 4, 1})$ (coupled to the
dynamical $({\bf 1, 4, 3})$) as a condition of the presence of ${\cal N}=4$ supersymmetry in the relevant system.

\section{Coupling to chiral superfields}
As was shown in \cite{SidSolo,IS2021}, the chiral linear multiplet ${\bf (2, 4, 2)}$ can be easily coupled to the linear mirror multiplets ${\bf (4, 4, 0)}$
and ${\bf (3, 4, 1)}$, in such a way that the interaction terms are constructed as superpotentials in chiral superspace. This proved possible because these multiplets
are described by (anti)chiral superfields. In this Section we show that the non-linear multiplet ${\bf (3, 4, 1)}$ also admits a description through chiral superfields,
which allows us  to couple it to the linear chiral multiplet ${\bf (2, 4, 2)}$. 

\subsection{Chiral solution}
To deal with chiral superfields it is convenient to pass to the notations
\bea
    &&\theta_{i}:=\theta_{i1}\,,\qquad
    \bar{\theta}^{i}:=\theta^i_2\,,\qquad
    \overline{\left(\theta_{i}\right)}=\bar{\theta}^{i},\nn
    &&D^{i}:=D^{i1},\qquad
    \bar{D}_i:=D^{2}_{i}\,,\qquad
    \overline{\left(D^i\right)} = \bar{D}_i\,,
\eea
that yields
\bea
    &&\delta \theta_{i}= \epsilon_{i}\,, \qquad \delta \bar{\theta}^{j}= \bar{\epsilon}^{j}, \qquad  \delta t = i\left(\bar{\epsilon}^{k}\theta_{k}+\epsilon_{k}\bar{\theta}^{k}\right),\qquad \overline{\left(\epsilon_{i}\right)}=\bar{\epsilon}^{i},\nn
    &&D^{i}=\frac{\partial}{\partial \theta_i}-i\,\bar{\theta}^i\partial_t\,,\qquad \bar{D}_i= -\,\frac{\partial}{\partial \bar{\theta}^i}+i\,\theta_i\partial_t\,.
\eea
One can split the triplet $V^{\alpha\beta}$ into real and complex superfields as $V_{12}=-\,i\,X$, $V_{11}=\sqrt{2}\,i\,\Phi$, $V_{22}=-\,\sqrt{2}\,i\,\bar{\Phi}$.
The non-linear constraints \eqref{NL341} are rewritten as
\bea
    &&\bar{D}_i\Phi = \frac{\sqrt{2}\,\Phi D_i\Phi}{X+iR}\,,\qquad
    \bar{D}_iX=\frac{\left(X-iR\right)D_i\Phi}{\sqrt{2}\left(X+iR\right)}-\frac{\Phi\bar{D}_i\bar{\Phi}}{X-iR}\,,\nn
    &&D^i\bar{\Phi} = -\,\frac{\sqrt{2}\,\bar{\Phi}\bar{D}^i\bar{\Phi}}{X-iR}\,,\qquad D^iX=-\,\frac{\left(X+iR\right)\bar{D}^i\bar{\Phi}}{\sqrt{2}\left(X-iR\right)}-\frac{\bar{\Phi}D^i\Phi}{X+iR}\,.
\eea
The complex superfield $\Phi$ becomes a chiral superfield in the limit $R\to\infty$.

For what follows, it is convenient to deal with yet another form of the triplet $V^{ij}$ given by
\bea
     &&Y=\frac{iR^2}{4\sqrt{2}}\left(\frac{1}{N^{12}}+\frac{1}{N^{21}}\right)=\frac{iR}{4}\left(R^2+V^{2}\right)\left[\frac{1}{\left(V^{2}-R^2-2R\,V_{12}\right)}-\frac{1}{\left(V^{2}-R^2+2R\,V_{12}\right)}\right],\nn
     &&U=-\,\frac{iR\,N^{22}}{2\sqrt{2}\,N^{21}}=\frac{iR^2\,V^{22}}{\sqrt{2}\left(V^{2}-R^2+2R\,V_{12}\right)}\,,\nn
     &&\bar{U}=\frac{iR\,N^{11}}{2\sqrt{2}\,N^{12}}=\frac{-\,iR^2\,V^{11}}{\sqrt{2}\left(V^{2}-R^2-2R\,V_{12}\right)}\,.
\eea
From \eqref{NL341} we derive that the complex superfield $U$ is chiral:
\bea
    \bar{D}_iU\left(t_{\rm L}\,,\theta_i\right)=0,\qquad \bar{D}_i=-\,\frac{\partial}{\partial \bar{\theta}^i}\,,\qquad t_{\rm L}=t+i\,\bar{\theta}^k\theta_k\,.
\eea
The full set of non-linear superfield constraints become
\bea
    &&\bar{D}_iU =0,\qquad
    D^iU = 
    -\,\frac{\sqrt{2}}{R}\bar{D}^{i}\left[Y\left(R^2+8U\bar{U}-4Y^2\right)^{\frac{1}{2}}-2i\left(Y^2-U\bar{U}\right)\right],\nn
    &&D^i\bar{U}=0,\qquad
    \bar{D}_i\bar{U}=
    \frac{\sqrt{2}}{R}D_i\left[Y\left(R^2+8U\bar{U}-4Y^2\right)^{\frac{1}{2}}+2i\left(Y^2-U\bar{U}\right)\right],\nn
    \bar{D}_{i}Y&=&-\left[\frac{D_iU}{\sqrt{2}}+\frac{2i}{R}\,U\bar{D}_i\bar{U}\left(1-2i\,Y\left(R^2+8U\bar{U}-4Y^2\right)^{-\frac{1}{2}}\right)\right]\nn
    &&\times\left[\frac{1}{R}\left(R^2+8U\bar{U}-4Y^2\right)^{-\frac{1}{2}}\left(R^2+8U\bar{U}-8Y^2\right)-\frac{4i}{R}\,Y\right]^{-1},\nn
    D^{i}Y&=& \left[\frac{\bar{D}^i\bar{U}}{\sqrt{2}}+\frac{2i}{R}\,\bar{U}D^iU\left(1+2i\,Y\left(R^2+8U\bar{U}-4Y^2\right)^{-\frac{1}{2}}\right)\right]\nn
    &&\times\left[\frac{1}{R}\left(R^2+8U\bar{U}-4Y^2\right)^{-\frac{1}{2}}\left(R^2+8U\bar{U}-8Y^2\right)+\frac{4i}{R}\,Y\right]^{-1}.\label{NL341_2}
\eea
The ${\rm SU}(2)_{\rm ext.}$ translations \eqref{S3tr_ext} act as
\bea
    &&\delta Y=\frac{2i}{R^2}\left(\bar{a}\,U-a\,\bar{U}\right)\left(R^2+8U\bar{U}-4Y^2\right)^{\frac{1}{2}}+\frac{4Y}{R^2}\left(\bar{a}\,U+a\,\bar{U}\right)
    -\frac{\alpha}{4R}\left(R^2+8U\bar{U}-4Y^2\right)^{\frac{1}{2}},\nn
    &&\delta U=a+\frac{i}{R}\,\alpha\,U+\frac{8\,\bar{a}\,U^2}{R^2}\,,\qquad
    \delta \bar{U}=\bar{a}-\frac{i}{R}\,\alpha\,\bar{U}+\frac{8\,a\,\bar{U}^2}{R^2}\,.\label{tr_trans}
\eea
The ${\rm SU}(2)_{\rm rot.}$ rotations \eqref{S3tr_diag} are rewritten as
\bea
    &&\delta Y=
    \frac{\sqrt{2}\left(R^2+8U\bar{U}-8Y^2\right)}{R\left(R^2+8U\bar{U}\right)}\left[i\left(\bar{b}\,U-b\,\bar{U}\right)\left(R^2+8U\bar{U}-4Y^2\right)^{\frac{1}{2}}-2Y\left(\bar{b}\,U+b\,\bar{U}\right)\right],\nn
    &&\delta U=\frac{\sqrt{2}\,b}{R}\left[2Y^2-2U\bar{U}+i\,Y\left(R^2+8U\bar{U}-4Y^2\right)^{\frac{1}{2}}\right]+\frac{2\sqrt{2}\,\bar{b}\,U^2}{R}-2i\beta\,U,\nn
    &&\delta \bar{U}=\frac{\sqrt{2}\,\bar{b}}{R}\left[2Y^2-2U\bar{U}-i\,Y\left(R^2+8U\bar{U}-4Y^2\right)^{\frac{1}{2}}\right] +\frac{2\sqrt{2}\,b\,\bar{U}^2}{R}+2i\beta\,\bar{U},\nn
    &&\delta D^{i}=i\bar{b}\,\bar{D}^i+i\beta\,D^{i},\qquad
    \delta \bar{D}_i=ib\,D_{i}-i\beta\,\bar{D}_i\,.\label{tr_rot}
\eea
It is straightforward to check that the constraints \eqref{NL341_2} are invariant under these translations and rotations.

The chiral superfield $U$ obtained as a solution of eqs. \eqref{NL341_2} is given by the expression
\bea
    U\left(t_{\rm L}\,,\theta_i\right) &=& u +\sqrt{2}\,\theta_k\psi^k - \frac{1}{2\sqrt{2}}\,\theta_k\theta^k\,C\left[\left(1 +\frac{8u\bar{u}}{R^2}- \frac{8y^2}{R^2}\right)-\frac{4iy}{R}\left(1 +\frac{8u\bar{u}}{R^2}- \frac{4y^2}{R^2}\right)^{\frac{1}{2}}\right]\nn
    &&-\,\frac{i}{\sqrt{2}}\,\theta_k\theta^k\,\dot{y}\left[\left(1 +\frac{8u\bar{u}}{R^2}- \frac{4y^2}{R^2}\right)^{-\frac{1}{2}}\left(1 +\frac{8u\bar{u}}{R^2}- \frac{8y^2}{R^2}\right)-\frac{4iy}{R}\right]\nn
    &&-\,\frac{2\sqrt{2}}{R}\,\theta_k\theta^k\,\dot{u}\bar{u}\left[\left(1 +\frac{8u\bar{u}}{R^2}- \frac{8y^2}{R^2}\right)-\frac{4iy}{R}\left(1 +\frac{8u\bar{u}}{R^2}- \frac{4y^2}{R^2}\right)^{\frac{1}{2}}\right]\left(1 +\frac{8u\bar{u}}{R^2}\right)^{-1}\nn
    &&+\,\frac{\sqrt{2}}{R}\,\theta_k\theta^k\left(u\dot{\bar{u}}+\dot{u}\bar{u}\right)\left[1-\frac{2iy}{R}\left(1 +\frac{8u\bar{u}}{R^2}- \frac{4y^2}{R^2}\right)^{-\frac{1}{2}}\right]+\theta_k\theta^k\left(\psi^2 {\rm\;term}\right).\label{U}
\eea
In this description, the field content ${\bf (3,4,1)}$ is identified with $\left(y,u,\bar{u},\psi^i,\bar{\psi}_j\,,C\right)$, where the last field $C$ coincides with the bosonic auxiliary field in the solution \eqref{V}.

In the chiral basis, the integration measure \eqref{D4} is rewritten  as
\bea
    d^4\theta = \frac{1}{4}\,D^{i}D_{i}\,\bar{D}_{j}\bar{D}^{j},
\eea
and the condition \eqref{quadra} as
\bea
    D^{(i}\bar{D}^{j)}L^{\alpha\beta}\left(Y,U,\bar{U}\right)=0.\label{DbarD}
\eea
Then the WZ action \eqref{WZaction} can be represented as:
\bea
    S_{\rm WZ}= i\int dt\left[D^{k}\bar{D}_{k}\,L^{12}\left(U,\bar{U},Y\right)+D^{i}D_{i}\,L^{22}\left(U,\bar{U},Y\right)-\bar{D}_{j}\bar{D}^{j}\,L^{11}\left(U,\bar{U},Y\right)\right].\label{WZL_2}
\eea
One can consider a special choice,
\bea
    L^{11}:=-\,\frac{i}{2}\,\bar{L}\left(\bar{U}\right),\qquad L^{22}:=\frac{i}{2}\,L\left(U\right),\qquad  L^{12}=0,
\eea
that corresponds to the superpotential
\be
    S_{\rm pot.} = \frac{1}{2}\int dt\left[D^{i}D_{i}\,L\left(U\right)+\bar{D}_{j}\bar{D}^{j}\,\bar{L}\left(\bar{U}\right)\right]=\int dt_{\rm L}\,d^2\theta\,L\left(U\right)+\int dt_{\rm R}\,d^2\bar{\theta}\,\bar{L}\left(\bar{U}\right).
\ee
Because of the (anti)chirality condition, the (anti)holomorphic functions $L\left(U\right)$ and $\bar{L}\left(\bar{U}\right)$ satisfy the quadratic condition \eqref{DbarD}.

\subsection{From the plane to a squashed sphere}
In \cite{IS2021} we considered the simplest WZ Lagrangian for the linear multiplet ${\bf (3, 4, 1)}$ given by
\bea
    S_{\rm plane} = \int dt\,D^{k}\bar{D}_{k}\left[\frac{1}{4}\left(Y^2-U\bar{U}\right)-\frac{c\,Y}{2}\right],\label{PlaneAction}
\eea
where the linear constraints are
\bea
    D^i\bar{U}=0,\qquad \bar{D}_i U=0,\qquad\sqrt{2}\,D_i Y =\bar{D}_i\bar{U},\qquad \sqrt{2}\,\bar{D}_i Y =-\,D_i U. \label{linear}
\eea
The real constant parameter $c$ corresponds to the FI term. The component Lagrangian reads
\bea
   {\cal L}_{\rm plane}=\frac{C}{2}\left(c-y\right)+\frac{i}{2}\left(u\dot{\bar{u}}-\dot{u}\bar{u}\right) -\frac{1}{4}\,\chi^{i}_{1}\chi_{i2}\,,\label{PlaneL}
\eea
where the contribution of the FI term proportional to the parameter $c$ reads
\bea
    {\cal L}_{\rm FI} = \frac{C}{2}\,c\,,\label{FILc}
\eea
and it is ${\cal N}=4$ invariant on its own.
The constraint $y\approx c$ defines a non-commutative plane\footnote{
The plane constraint can be obtained from the 2-sphere equation \eqref{S2inR3} in the planar limit $r\to\infty$. The limit for both linear and non-linear cases is presented in Appendix \ref{AppB}.}.

Here we consider the simplest modification of the non-commutative plane action \eqref{PlaneAction} for the non-linear multiplet \eqref{NL341_2} that leads to a squashed 2-sphere.
The action is composed of the two parts
\bea
    S_{\rm sq.sphere} = \int dt\,D^{k}\bar{D}_{k}\left[\frac{1}{4}\left(Y^2-U\bar{U}\right)-\frac{c\,Y}{2R}\left(R^2+8U\bar{U}-4Y^2\right)^{\frac{1}{2}}\right].\label{sq.sphereA}
\eea
For both parts we can prove that
\bea
    D^{(i}\bar{D}^{j)}\left(Y^2-U\bar{U}\right)=0,\qquad D^{(i}\bar{D}^{j)}\left[Y\left(R^2+8U\bar{U}-4Y^2\right)^{\frac{1}{2}}\right]=0\,,
\eea
so they are  ${\cal N}=4$ supersymmetric separately. The action can be brought to the superpotential form:
\bea
    S_{\rm sq.sphere} = \frac{iR}{16\sqrt{2}}\int dt\left[\left(1+\frac{4ic}{R}\right)D^{i}D_{i}U-\left(1-\frac{4ic}{R}\right)\bar{D}_{j}\bar{D}^{j}\bar{U}\right].\label{sq.sphere}
\eea
It is invariant under the ${\rm U}(1)_{\rm rot.}$ rotation with the parameter $\beta$ from \eqref{tr_rot}.
One could consider any other ${\rm U}(1)_{\rm rot.}$ - invariant generalization of \eqref{PlaneAction} written in the form \eqref{WZL_2},
but we stick to the simplest  choice given by \eqref{sq.sphere}.

The component Lagrangian can be  written, up to a total time derivative, as
\bea
    {\cal L}_{\rm sq.sphere}&=&\frac{i}{2}\left(u\dot{\bar{u}}-\dot{u}\bar{u}\right)\left[\left(1 +\frac{8u\bar{u}}{R^2}- \frac{8y^2}{R^2}\right)+\frac{16\,c\,y}{R^2}
    \left(1 +\frac{8u\bar{u}}{R^2}- \frac{4y^2}{R^2}\right)^{\frac{1}{2}}\right]\left(1 +\frac{8u\bar{u}}{R^2}\right)^{-1}\nn
    &&+\,\frac{\tilde{C}}{2}\left[c\left(1 +\frac{8u\bar{u}}{R^2}- \frac{8y^2}{R^2}\right)-y\left(1 +\frac{8u\bar{u}}{R^2}- \frac{4y^2}{R^2}\right)^{\frac{1}{2}}\right]+\psi^2 {\rm\;term}, \label{sq.sphereL}
\eea
where
 $\tilde{C}$ is the redefined auxiliary field
\bea
    \tilde{C}=C+\frac{R}{2}\,\partial_{t}\ln{\left(1 +\frac{8u\bar{u}}{R^2}\right)}.
\eea
We use the same notation for the component fields as in the case of linear multiplet ${\bf (3, 4, 1)}$, in order to visualize the relation between both spin multiplets. The limit $R\to\infty$ leads to \eqref{PlaneL}.

\paragraph{Remark.}
After a proper redefinition of the component fields (see Appendix C), the Lagrangian \eqref{sq.sphereL} can be brought to the form
\be
    {\cal L}_{\rm sq.sphere}=\frac{\hat{C}}{2}\left(c-\hat{y}\right)+\frac{i}{2}\left(\hat{u}\dot{\bar{\hat{u}}}-\dot{\hat{u}}\bar{\hat{u}}\right)
    -\frac{1}{4}\,\hat{\chi}^{k}\bar{\hat{\chi}}_{k}+\frac{1}{R}\left[\hat{\chi}_{k}\hat{\chi}^{k}\,A\left(\hat{y},\hat{u},\bar{\hat{u}}\right)+\bar{\hat{\chi}}^{k}\bar{\hat{\chi}}_{k}\bar{A}
    \left(\hat{y},\hat{u},\bar{\hat{u}}\right)\right],
\ee
where $A\left(\hat{y},\hat{u},\bar{\hat{u}}\right)$ is non-singular in the limit $R\to \infty$. One sees that the bosonic part takes the same form as  in the plane case. However, the fermionic part {\it cannot} be
simplified to $\sim\hat{\chi}^{k}\bar{\hat{\chi}}_{k}$\,, {\it i.e.} the off-shell Lagrangian ${\cal L}_{\rm sq.sphere}$ is not equivalent to \eqref{PlaneL}\footnote{On shell, fermionic fields are zero, so we come back to
the plane case. However, off shell and when interactions with other multiplets are switched on, elimination of the auxiliary fermionic fields yields different results for linear and non-linear spin multiplets.
It is worthwhile to  note an analogy with ${\cal N}=2$, $4D$ Einstein supergravity theories (for which $4D$ analogs of the linear and non-linear $d=1$ multiplets serve as conformal compensators, see, e.g.,\cite{Kuz2}
and refs. therein): when interactions with matter multiplets are absent, these theories are on-shell equivalent, but the equivalence is lost in the presence of matter couplings.}.  \\

The constraint imposed by the equation of motion for $\tilde{C}$ reads
\bea
    c\left(1+\frac{8u\bar{u}}{R^2}-\frac{8y^2}{R^2}\right)-y\left(1+\frac{8u\bar{u}}{R^2}-\frac{4y^2}{R^2}\right)^{\frac{1}{2}}\approx 0.\label{C}
\eea
To clarify the meaning of this constraint, let us pass to the coordinates \eqref{x4} in $\mathbb{R}^4$ given by
\bea
     &&x_{1}=\sqrt{2}\,i\left[\frac{u}{\left(1+\frac{8u\bar{u}}{R^2}-\frac{4y^2}{R^2}\right)^{\frac{1}{2}}-\frac{2iy}{R}}-\frac{\bar{u}}{\left(1+\frac{8u\bar{u}}{R^2}-\frac{4y^2}{R^2}\right)^{\frac{1}{2}}+\frac{2iy}{R}}\right],\nn
     &&x_{2}=-\,\sqrt{2}\left[\frac{u}{\left(1+\frac{8u\bar{u}}{R^2}-\frac{4y^2}{R^2}\right)^{\frac{1}{2}}-\frac{2iy}{R}}+\frac{\bar{u}}{\left(1+\frac{8u\bar{u}}{R^2}-\frac{4y^2}{R^2}\right)^{\frac{1}{2}}+\frac{2iy}{R}}\right],\nn
     &&|x_{3}|=R\left(1+\frac{8u\bar{u}}{R^2}-\frac{4y^2}{R^2}\right)^{\frac{1}{2}}\left(1+\frac{8u\bar{u}}{R^2}\right)^{-1},\qquad
x_{4}=2y\left(1+\frac{8u\bar{u}}{R^2}\right)^{-1}.
\eea
Note that the absolute value of $y$ is limited  to the range $0\leqslant 2\,|y|\leqslant \left(R^2+8u\bar{u}\right)^{\frac{1}{2}}$.
The constraint \eqref{C} is rewritten as
\bea
     \frac{R^2}{\left[\left(x_3\right)^2+\left(x_4\right)^2\right]^2}\left[c\left(x_3\right)^2-c\left(x_4\right)^2-\frac{R}{2}\,x_4\,|x_3|\right]\approx 0,\label{C2}
\eea
and it has the two plane solutions:
\bea
     {\rm 1)}\quad x_4\approx \frac{4\,c\,|x_3|}{\sqrt{R^2+16\,c^2}+R}\,,\qquad
     {\rm 2)}\quad x_4\approx -\,\frac{4\,c\,|x_3|}{\sqrt{R^2+16\,c^2}-R}\,.
\eea
Without loss of generality, we can choose the first option since it yields the proper limit as $R\to \infty$, {\it i.e.} $y\approx c$.
The constraint \eqref{C2} can be interpreted as a section of the sphere $S^3$ by a plane and it defines a squashed 2-sphere:
\bea
     {\rm 1)}\quad R^2-\left(x_{1}\right)^2-\left(x_{2}\right)^2-\left(x_{3}\right)^2\approx\frac{16\,c^2\left(x_{3}\right)^2}{\left(\sqrt{R^2+16\,c^2}+R\right)^2} \quad\Rightarrow \qquad\qquad\qquad\qquad\qquad\qquad\nn
     \Rightarrow\quad \left(x_{1}\right)^2+\left(x_{2}\right)^2+ \frac{2\sqrt{1+\frac{16\,c^2}{R^2}}}{1 + \sqrt{1+\frac{16\,c^2}{R^2}}}\left(x_{3}\right)^2 \approx R^2.
\eea
It respects only some ${\rm U}(1)$ symmetry from ${\rm SO}(4)$, {\it viz.}, the one that keeps invariant the bilinear form $\left(x_{1}\right)^2+\left(x_{2}\right)^2$.
For $c=0$ the 2-sphere becomes round, but the constraint \eqref{C2} still respects only ${\rm U}(1)$ symmetry.

Let us present the relevant Dirac brackets (with $c=0$ for simplicity):
\bea
    &&\left\lbrace u, \bar{u}\right\rbrace_{\rm DB} = i\left(1 +\frac{8u\bar{u}}{R^2}- \frac{8y^2}{R^2}\right)\left(1 +\frac{8u\bar{u}}{R^2}\right)^2\left[\left(1 +\frac{8u\bar{u}}{R^2}- \frac{8y^2}{R^2}\right)^2+\frac{8u\bar{u}}{R^2}\left(1 +\frac{8u\bar{u}}{R^2}\right)^2\right]^{-1},\nn
    &&\left\lbrace u, y\right\rbrace_{\rm DB} = -\,\frac{4iyu}{R^2}\left(1 +\frac{8u\bar{u}}{R^2}\right)^2\left[\left(1 +\frac{8u\bar{u}}{R^2}- \frac{8y^2}{R^2}\right)^2+\frac{8u\bar{u}}{R^2}\left(1 +\frac{8u\bar{u}}{R^2}\right)^2\right]^{-1},\nn
    &&\left\lbrace \bar{u}, y\right\rbrace_{\rm DB} = \frac{4iy\bar{u}}{R^2}\left(1 +\frac{8u\bar{u}}{R^2}\right)^2\left[\left(1 +\frac{8u\bar{u}}{R^2}- \frac{8y^2}{R^2}\right)^2+\frac{8u\bar{u}}{R^2}\left(1 +\frac{8u\bar{u}}{R^2}\right)^2\right]^{-1}.
\eea
The plane bracket $\left\lbrace u, \bar{u}\right\rbrace_{\rm DB} = i$ \, is reproduced in the limit $R\to \infty$.

Finally, it is worth noting that the part proportional to the parameter $c$ in \eqref{sq.sphereL} involves two independent ${\cal N}=4$ supersymmetric terms. One of them is the invariant FI term,
\bea
    {\cal L}_{\rm FI} = \frac{\tilde{C}}{2}\,c,
\eea
because $\tilde{C}$ transforms into a full time derivative under ${\cal N}=4$ supersymmetry. In the limit $R\to\infty$, it comes to the FI term \eqref{FILc}, while the rest of  the contribution proportional
to $c$ disappears.  One can consider a more general two-parameter deformation of $S^2$ by adding to \eqref{sq.sphereL} an independent ${\cal N}=4$ supersymmetric FI term,
\bea
(3.22) \;\; \Rightarrow \;\; (3.22) + \frac{\tilde{C}}{2}\,\gamma\,, \qquad \gamma = {\rm const}\,,
\eea
so that the embedding constraint \eqref{C} is modified as
\bea
   \gamma +  c\left(1+\frac{8u\bar{u}}{R^2}-\frac{8y^2}{R^2}\right)-y\left(1+\frac{8u\bar{u}}{R^2}-\frac{4y^2}{R^2}\right)^{\frac{1}{2}}\approx 0.\label{C1}
\eea
It is rather difficult to solve because it generally amounts to a fourth-order algebraic equation for the $S^3$ coordinate $x_4$. Even for the simple choice $c=0, \gamma\neq 0$, the analog of \eqref{C2} is
\bea
\gamma\,\left[\left(x_3\right)^2+\left(x_4\right)^2\right]^2 - \frac{R^3}{2}\,|x_3|x_4 \approx 0\,.
\eea

\subsection{Interaction}
We proceed to the coupling of the linear chiral multiplet ${\bf (2, 4, 2)}$ to the non-linear multiplet ${\bf (3, 4, 1)}$. The linear multiplet ${\bf (2, 4, 2)}$ is described
by the chiral superfield $Z$:
\bea
    \bar{D}_{i}Z=0,\qquad Z\left(t_{\rm L},\theta_i\right)=z+\sqrt{2}\,\theta_{k}\xi^k+\theta_{k}\theta^k B.\label{Z242}
\eea
We consider the multiplet ${\bf (2, 4, 2)}$ as dynamical, so its action must include a kinetic part and may also involve a superpotential part:
\bea
    S_{\bf (2,4,2)} = \frac{1}{4}\int dt\,d\theta^2\,d^2\bar{\theta}\,K\left(Z,\bar{Z}\right)+\frac{1}{2}\int dt_{\rm L}\,d^2\theta\,{\cal K}\left(Z\right)
    +\frac{1}{2}\int dt_{\rm R}\,d^2\bar{\theta}\,\bar{\cal K}\left(\bar{Z}\right).\label{242L}
\eea
The interaction of both multiplets is ensured by the following superpotential:
\bea
    S_{\rm int.}=\frac{\mu}{2}\int dt_{\rm L}\,d^2\theta\,{\cal F}\left(U,Z\right)+\frac{\mu}{2}\int dt_{\rm R}\,d^2\bar{\theta}\,\bar{\cal F}\left(\bar{U},\bar{Z}\right).\label{intL}
\eea
The total action is a sum of \eqref{242L}, \eqref{intL} and \eqref{WZaction}.

Let us consider the squashed sphere action \eqref{sq.sphere} coupled to the kinetic term $K\left(Z,\bar{Z}\right)$ with the simplest choice of the interaction as
\bea
{\cal F}\left(U,Z\right)=U\,Z. \label{simplestUZ}
\eea
We obtain
\bea
    S &=& \frac{1}{4}\int dt\,d\theta^2\,d^2\bar{\theta}\,K\left(Z,\bar{Z}\right)+\frac{iR}{8\sqrt{2}}\,\int dt_{\rm L}\,d^2\theta\left(1+\frac{4ic}{R}-\frac{4\sqrt{2}\,i\mu}{R}\,Z\right)U\nn
    &&-\,\frac{iR}{8\sqrt{2}}\,\int dt_{\rm R}\,d^2\bar{\theta}\left(1-\frac{4ic}{R}+\frac{4\sqrt{2}\,i\mu}{R}\,\bar{Z}\right)\bar{U}.
\eea
We will limit our consideration to the bosonic Lagrangian which is given by (up to total time derivative)
\bea
    {\cal L}\left.\right|_{\rm bos.}&=&\left(\dot{\bar{z}}\dot{z} + \bar{B}B\right)\partial_{z}\partial_{\bar{z}}K\left(z,\bar{z}\right)+\mu\left(B\,u+\bar{B}\,\bar{u}\right)\nn
    &&+\,\frac{\tilde{C}}{2}\left\lbrace\left[c-\frac{\mu}{\sqrt{2}}\left(z+\bar{z}\right)\right]\left(1 +\frac{8u\bar{u}}{R^2}- \frac{8y^2}{R^2}\right)-y\left[1-\frac{2\sqrt{2}\,i\mu}{R}\left(z-\bar{z}\right)\right]\left(1 +\frac{8u\bar{u}}{R^2}- \frac{4y^2}{R^2}\right)^{\frac{1}{2}}\right\rbrace\nn
    &&+\,\frac{i}{2}\left(u\dot{\bar{u}}-\dot{u}\bar{u}\right)\left[1-\frac{2\sqrt{2}\,i\mu}{R}\left(z-\bar{z}\right)\right]\left(1 +\frac{8u\bar{u}}{R^2}- \frac{8y^2}{R^2}\right)\left(1 +\frac{8u\bar{u}}{R^2}\right)^{-1}\nn
    &&+\,\frac{8iy\left(u\dot{\bar{u}}-\dot{u}\bar{u}\right)}{R^2}\left[c-\frac{\mu}{\sqrt{2}}\left(z+\bar{z}\right)\right]\left(1 +\frac{8u\bar{u}}{R^2}- \frac{4y^2}{R^2}\right)^{\frac{1}{2}}\left(1 +\frac{8u\bar{u}}{R^2}\right)^{-1}\nn
    &&-\,\frac{\sqrt{2}\,\mu }{8}\left[R\left(1 +\frac{8u\bar{u}}{R^2}- \frac{8y^2}{R^2}\right)\left(\dot{z}+\dot{\bar{z}}\right)-4iy\left(1 +\frac{8u\bar{u}}{R^2}
    - \frac{4y^2}{R^2}\right)^{\frac{1}{2}}\left(\dot{z}-\dot{\bar{z}}\right)\right].\label{totalL}
\eea
The equation of motion for $\tilde{C}$ imposes the constraint
\be
    \left[c-\frac{\mu}{\sqrt{2}}\left(z+\bar{z}\right)\right]\left(1 +\frac{8u\bar{u}}{R^2}- \frac{8y^2}{R^2}\right)-y\left[1-\frac{2\sqrt{2}\,i\mu}{R}\left(z-\bar{z}\right)\right]\left(1 +\frac{8u\bar{u}}{R^2}- \frac{4y^2}{R^2}\right)^{\frac{1}{2}}\approx 0.
\ee
We observe that it looks the same as the constraint \eqref{C}, though with the modified squashing parameter $\tilde{c}$
depending on the coordinates $z$ and $\bar{z}$:
\be
\tilde{c}\left(z,\bar{z}\right)=\left[c-\frac{\mu}{\sqrt{2}}\left(z+\bar{z}\right)\right]\left[1-\frac{2\sqrt{2}\,i\mu}{R}\left(z-\bar{z}\right)\right]^{-1}.
\ee
Therefore, the squashed 2-sphere is specified by the parameter $\tilde{c}$ ``gauged'' by points of
the underlying K\"ahler manifold. The dependence on $z, \bar z$ can become more involved, when switching on  less trivial interactions than the simplest one \eqref{simplestUZ}.

\section{Summary and outlook}
We considered the non-linear version of the mirror multiplet ${\bf (3, 4, 1)}$ as a semi-dynamical multiplet and constructed its action \eqref{WZaction}.
We showed that the quadratic constraint \eqref{quadra} for the superfield function induces the Laplace-Beltrami equation \eqref{LBeq} on the sphere $S^3$.
We presented the ${\rm SU}(2)_{\rm rot.}$ invariant action \eqref{SU2} that describes the
embedding of the 2-dimensional sphere $S^2$ into $S^3$. The limit $R\to\infty$ leads to the  embedding of $S^2$ into $\mathbb{R}^3$.
As a simplest generalization of the non-commutative plane we presented the action \eqref{sq.sphereA} that describes the squashed sphere $S^2$ embedded into $S^3$.
We coupled this squashed sphere to the dynamical mirror chiral multiplet ${\bf (2, 4, 2)}$ and constructed their interaction as a superpotential. An interesting feature of this interaction
is that it induces a dependence on the complex bosonic field $z, \bar z$ of the dynamical multiplet in the squashing parameter, $c \rightarrow \tilde{c}(z, \bar  z)$.
Models with the linear multiplet ${\bf (3, 4, 1)}$ are reproduced in the limit $R\to\infty$.

It is yet unclear how to couple the non-linear spin multiplet to other dynamical ${\cal N}=4$ multiplets, in particular to the multiplet  ${\bf (1, 4, 3)}$.
Perhaps, this can be accomplished within the harmonic superspace description \cite{HSSd1,FIL2012}.

Let us specify  some further directions of study. It would be interesting to find the torus solution $S^{1} \times S^{1}$ as an embedding in $S^3$.
One more  tempting problem is to consider the non-linear multiplet ${\bf (4,4,0)}$ \cite{DI2011} as semi-dynamical.
There is also some interest in treating the non-linear multiplet ${\bf (3, 4, 1)}$ as the dynamical one and constructing the relevant $d=1$ sigma-model type actions.
For example, a model of $3D$ supersymmetric particle
in the monopole background \cite{IKS2,BKS2010} could be generalized to the modified potentials \eqref{UA}.

We showed that the non-linear constraints \eqref{NL341} are invariant under the superfield reflection $V^{\alpha\beta}\to -\,R^2\,V^{\alpha\beta}/V^2$.
It would be interesting to find out  possible implications of this discrete transformation in the sigma-model and WZ actions of the non-linear multiplet.

\section*{Acknowledgements}
The authors thank Sergey Fedoruk and Gor Sarkissian for useful discussions and comments. The research
was supported by the Russian Science Foundation Grant No 21-12-00129.

\appendix

\section{Geometry of the 3-sphere}\label{AppA}
The 3-dimensional metric defined by \eqref{metric} reads
\bea
    ds^2 = g_{(\alpha\beta)\,(\lambda\mu)}\,dv^{\alpha\beta}dv^{\lambda\mu}\,,\qquad g_{(\alpha\beta)\,(\lambda\mu)}=-\,\varepsilon_{\lambda(\alpha}\varepsilon_{\beta)\mu}\left(1+\frac{v^2}{R^2}\right)^{-2}.
\eea
Its inverse is
\bea
    g^{(\lambda\mu)\,(\gamma\delta)}=-\,\varepsilon^{\gamma(\lambda}\varepsilon^{\mu)\delta}\left(1+\frac{v^2}{R^2}\right)^{2},\qquad g_{(\alpha\beta)\,(\lambda\mu)}\,g^{(\lambda\mu)\,(\gamma\delta)}=\delta^{\gamma}_{(\alpha}\delta^{\delta}_{\beta)}\,.
\eea
The explicit form of Laplace-Beltrami equation \eqref{LBeq} is then as follows
\bea
    {\bf \Delta}_{S^3}=\frac{1}{\sqrt{|g|}}\,\partial_{\alpha\beta}\left(\sqrt{|g|}\,g^{(\alpha\beta)\,(\lambda\mu)}\right)\partial_{\lambda\mu} = \left(1+\frac{v^2}{R^2}\right)^{2}\left[\partial^{\alpha\beta}-\frac{v^{\alpha\beta}}{R^2}\left(1+\frac{v^2}{R^2}\right)^{-1}\right]\partial_{\alpha\beta}\,.
\eea
The relevant Christoffel coefficients read
\bea
    &&\nabla_{\alpha\beta}\,g_{(\lambda\mu)\,(\gamma\delta)}=\partial_{\alpha\beta}\,g_{(\lambda\mu)\,(\gamma\delta)}-\Gamma^{(\sigma\rho)}_{(\alpha\beta)\,(\lambda\mu)}\,g_{(\sigma\rho)\,(\gamma\delta)}-\Gamma^{(\sigma\rho)}_{(\alpha\beta)\,(\gamma\delta)}\,g_{(\lambda\mu)\,(\sigma\rho)}=0,\nn
    &&\Gamma^{(\sigma\rho)}_{(\alpha\beta)\,(\lambda\mu)}=-\,\frac{1}{R^2}\left(1+\frac{v^2}{R^2}\right)^{-1}\left(\delta^{\sigma}_{(\alpha}\delta^{\rho}_{\beta)}\,v_{\lambda\mu}+\delta^{\sigma}_{(\lambda}\delta^{\rho}_{\mu)}\,v_{\alpha\beta}+\varepsilon_{\lambda(\alpha}\varepsilon_{\beta)\mu}\,v^{\sigma\rho}\right).
\eea
The Laplace-Beltrami equation \eqref{LBU} for the scalar ${\cal U}$ can be then rewritten as
\bea
    {\bf \Delta}_{S^3}\,{\cal U}=g^{(\alpha\beta)\,(\lambda\mu)}\,\nabla_{\lambda\mu}\,\nabla_{\alpha\beta}\,{\cal U}=\left(1+\frac{v^2}{R^2}\right)^{2}\nabla^{\alpha\beta}\,\nabla_{\alpha\beta}\,{\cal U}=0, \quad \nabla^{\alpha\beta} \equiv
    \varepsilon^{\alpha\rho}\varepsilon^{\beta\gamma}\nabla_{\rho\gamma}\,.
\eea
The explicit form of abbreviations in eqs. \eqref{rotA} and \eqref{divA} is
\bea
    &&{\rm\bf rot}_{S^3}\,{\cal A} = |g|^{-1/2}\,\varepsilon^{\mu\lambda}\,\nabla_{\alpha(\lambda}\,{\cal A}_{\mu)\beta}\,,\qquad {\rm\bf grad}_{S^3}\,{\cal U}
    =g^{(\alpha\beta)\,(\lambda\mu)}\,\nabla_{\lambda\mu}\,{\cal U}, \nn
    &&{\rm\bf div}_{S^3}\,{\cal A} = g^{(\alpha\beta)\,(\lambda\mu)}\,\nabla_{\lambda\mu}\,{\cal A}_{\alpha\beta}\,.
\eea

\section{Planar limit}\label{AppB}
As was shown in \cite{IS2021}, the non-commutative
plane is the planar limit of the fuzzy sphere $S^2$.  The equation of the 2-sphere \eqref{S2inR3} embedded in $\mathbb{R}^3$ is rewritten up to a constant factor as
\bea
    \frac{1}{2}\left[r-\frac{r^2}{\sqrt{\left(y-c-r\right)^2+2u\bar{u}}}\right]\approx 0,\quad v_{12}=-\,i\left(y-c-r\right),\quad v_{11}=\sqrt{2}\,i\,u,\quad v_{22}=-\,\sqrt{2}\,i\,\bar{u}.
\eea
Then the limit $r\to\infty$ yields the constraint for the plane \eqref{PlaneL}: $\left(c-y\right)/2\approx 0$.

In the same way, the planar limit can be taken for the 2-sphere embedded in $S^3$ according to the constraint \eqref{S2inS3}.
To accomplish this in an unambiguous way, we make the particular choice $r=R$, {\it i.e.} relate the planar limit $r\to\infty$ to the limit $R\to\infty$. The suitable embedding
constraint is obtained by multiplying \eqref{S2inS3} by $R^2/2$ and is expressed as
\bea
    &&\frac{1}{4}\left[\sqrt{\left(y-c-r\right)^2+2u\bar{u}}-\frac{r^2}{\sqrt{\left(y-c-r\right)^2+2u\bar{u}}}\right]\approx 0,\nn
    &&v_{12}=-\,i\left(y-c-r\right),\quad v_{11}=\sqrt{2}\,i\,u,\quad v_{22}=-\,\sqrt{2}\,i\,\bar{u}.
\eea
In the limit $r\to\infty$, we again find $\left(c-y\right)/2\approx 0$.

\section{Rescalings}\label{AppC}
We start with the WZ Lagrangian \eqref{sq.sphereL} written in the form
\bea
    {\cal L}_{\rm sq.sphere} = \frac{i}{2}\left(u\dot{\bar{u}} - \dot{u}\bar{u}\right)\left(1 - \frac{8y^2}{R^2 A}\right)\left(1 + \frac{16\,c\,\hat{y}}{R^2}\right) +
\frac{\hat{C}}{2}\left(c - \hat{y}\right)+\psi^2 {\rm\;term}, \label{I}
\eea 
where
\bea
    A:=1+\frac{8u\bar{u}}{R^2},\qquad \hat{y} := y \left(A-\frac{4y^2}{R^2}\right)^{\frac{1}{2}}\left(A-\frac{8y^2}{R^2}\right)^{-1},\qquad \hat{C}:=\tilde{C}\left(A - \frac{8y^2}{R^2}\right).\nonumber
\eea
We further define
\bea
1 - \frac{8y^2}{R^2A} = X \quad\rightarrow\quad  1-\frac{4y^2}{R^2A} = \frac{1}{2}\left(1 + X\right), \quad \frac{8y^2}{R^2A} = 1 -X,\nonumber
\eea
and obtain that
\bea
\frac{8\hat{y}^2}{R^2} = \frac{8y^2}{R^2A}\left(1-\frac{4y^2}{R^2A}\right)\left(1 - \frac{8y^2}{R^2A} \right)^{-2}= \frac{\left(1 + X\right)\left(1 - X\right)}{2X^2}
= \frac{1 - X^2}{2X^2} \,,\nonumber
\eea
whence
\be
X^2 =\left(1 + \frac{16\hat{y}^2}{R^2}\right)^{-1} \;\;\;\Rightarrow \;\;\; X = \pm\,\left(1 + \frac{16\hat{y}^2}{R^2}\right)^{-\frac{1}{2}}. \nonumber
\ee

For the first solution the first term in \eqref{I} is reduced to
\bea
    \frac{i}{2}\left(u\dot{\bar{u}} - \dot{u}\bar{u}\right)\left(1 + \frac{16\,c\,\hat{y}}{R^2}\right)\left(1 + \frac{16\hat{y}^2}{R^2}\right)^{-\frac{1}{2}}.\nonumber
\eea
The second solution for $X$ gives the equivalent result: the sign minus can be compensated by redefining
\bea
w =\bar{u}, \qquad \bar{w} = u.\nonumber
\eea
In both cases, the coefficients of $\left(u\dot{\bar{u}} - \dot{u}\bar{u}\right)$ can be brought to constants by the proper rescalings of $u$ and $\bar{u}$. As was already mentioned, the
complete equivalency with the planar model action cannot be achieved because the non-trivial dependence on the bosonic fields is retained in the fermionic terms.

\end{document}